\documentclass[sigconf]{acmart}
\usepackage{graphicx}
\usepackage{subfigure}
\usepackage{times}
\usepackage{soul}
\usepackage{url}
\usepackage[utf8]{inputenc}
\usepackage{caption}
\usepackage{amsmath}
\usepackage{amsthm}
\usepackage{booktabs}
\usepackage{algorithm}
\usepackage{algorithmic}
\usepackage{natbib}
\usepackage{amsfonts}
\usepackage[normalem]{ulem}
\usepackage[switch]{lineno}
\usepackage{multirow}
\usepackage{balance}
\AtBeginDocument{%
  \providecommand\BibTeX{{%
    \normalfont B\kern-0.5em{\scshape i\kern-0.25em b}\kern-0.8em\TeX}}}

\copyrightyear{2021} 
\acmYear{2021} 
\setcopyright{acmcopyright}
\acmConference[CIKM '21]{Proceedings of the 30th ACM International Conference on Information and Knowledge Management}{November 1--5, 2021}{Virtual Event, QLD, Australia}
\acmBooktitle{Proceedings of the 30th ACM International Conference on Information and Knowledge Management (CIKM '21), November 1--5, 2021, Virtual Event, QLD, Australia}
\acmPrice{15.00}
\acmDOI{10.1145/3459637.3482327}
\acmISBN{978-1-4503-8446-9/21/11}

\begin{document}
\title{Click-Through Rate Prediction with Multi-Modal Hypergraphs}

\author{Li He}
\authornote{Authors contributed equally to this research.}
\email{li.he-1@student.uts.edu.au}
\orcid{}
\affiliation{%
  \institution{University of Technology Sydney}
  \streetaddress{}
  \city{Sydney}
  \state{}
  \country{Australia}
  \postcode{}
}

\author{Hongxu Chen}
\authornote{Corresponding authors.}
\authornotemark[1]
\email{hongxu.chen@uts.edu.au}
\affiliation{%
  \institution{University of Technology Sydney}
  \streetaddress{}
  \city{Sydney}
  \state{}
  \country{Australia}
  \postcode{}
}

\author{Dingxian Wang}
\authornotemark[1]
\email{diwang@ebay.com}
\orcid{}
\affiliation{%
  \institution{eBay Research America}
  \streetaddress{}
  \city{Seattle}
  \state{}
  \country{United States}
  \postcode{}
}

\author{Jameel Shoaib}
\affiliation{%
  \institution{University of Essex}
  \streetaddress{}
  \city{Colchester}
  \country{United Kingdom}}
\email{shoaib.jameel@gmail.com}

\author{Philip Yu}
\affiliation{%
  \institution{University of Illinois at Chicago}
  \city{Chicago}
  \country{United States}}
 \email{psyu@uic.edu}

\author{Guandong Xu}
\authornotemark[2]
\affiliation{%
 \institution{University of Technology Sydney}
 \streetaddress{}
 \city{Sydney}
 \state{}
 \country{Australia}}
\email{guandong.xu@uts.edu.au}

\renewcommand{\shortauthors}{Li and Hongxu, et al.}

\begin{abstract}

Advertising is critical to many online e-commerce platforms such as e-Bay and Amazon. One of the important signals that these platforms rely upon is the click-through rate (CTR) prediction. The recent popularity of multi-modal sharing platforms such as TikTok has led to an increased interest in online micro-videos. It is, therefore, useful to consider micro-videos to help a merchant target micro-video advertising better and find users' favourites to enhance user experience. Existing works on CTR prediction largely exploit unimodal content to learn item representations. A relatively minimal effort has been made to leverage multi-modal information exchange among users and items. We propose a model to exploit the temporal user-item interactions to guide the representation learning with multi-modal features, and further predict the user click rate of the micro-video item. We design a Hypergraph Click-Through Rate prediction framework (HyperCTR) built upon the hyperedge notion of hypergraph neural networks, which can yield modal-specific representations of users and micro-videos to better capture user preferences. We construct a time-aware user-item bipartite network with multi-modal information and enrich the representation of each user and item with the generated interests-based user hypergraph and item hypergraph. Through extensive experiments on three public datasets, we demonstrate that our proposed model significantly outperforms various state-of-the-art methods.

\end{abstract}



\maketitle

\section{Introduction}
Click-Through Rate (CTR) prediction has become one of the core components of modern advertising on many e-commerce platforms. The goal is to predict customers' click probability on wide range of items. Existing works on CTR prediction only focus on modeling pairwise interactions from uni-modal features which might not lead to satisfactory results. This existing gap leads to new opportunities where we can exploit the widely available multi-modal features which is largely unexplored. Besides, they can give complementary information to the model which alone cannot be obtained via uni-modal modeling. AutoFIS~\cite{Liu2020AutoFIS} and UBR4CTR~\cite{Qin2020} are recent Factorization Machine (FM)~\cite{2010fm} based models with multi-layer perceptron (MLP) which mainly utilize user-item interactions features. To supplement the lack of additional information, deep neural networks (DNNs) are also explored with automated feature engineering. For example, DSTN~\cite{2019DeepSpatio} leverages DNNs-based method to fuse additional auxiliary data and item information to further uncover hidden information. Although these representative works have achieved good performance, there are still limited exploration on modeling multi-modal features and how they could contribute towards the model performance.

Recently, the wide-spreading influence of micro-video sharing platforms, e.g., Tiktok~\footnote{https://www.tiktok.com/} and Kuaishou~\footnote{https://www.kuaishou.com/} make them a popular platform for socialising, sharing and advertising as micro-videos. These videos are compact and come with rich multimedia content from multiple modalities, i.e., textual, visual, as well as acoustic information. Motivated by this, we propose a novel method which addresses the limitations in current methods and improve CTR prediction performance through micro-videos. However, modeling multi-modal features extracted from micro-videos for CTR prediction in a holistic way is not straightforward. First, in a typical setting of CTR prediction, the interactions between users and items are normally sparse, and the sparsity issue becomes even more severe (in magnitude of number of modalities) when taking into account multi-modal features. For example, compared to uni-modal feature space, the sparsity of a dataset is tripled when considering visual, acoustic and text features of a target item. Therefore, effectively mitigating the sparsity issues introduced by multi-modal features without compromising upon the performance of the model is the key to this problem.

\begin{figure}
  \centering
  \includegraphics[scale=0.25]{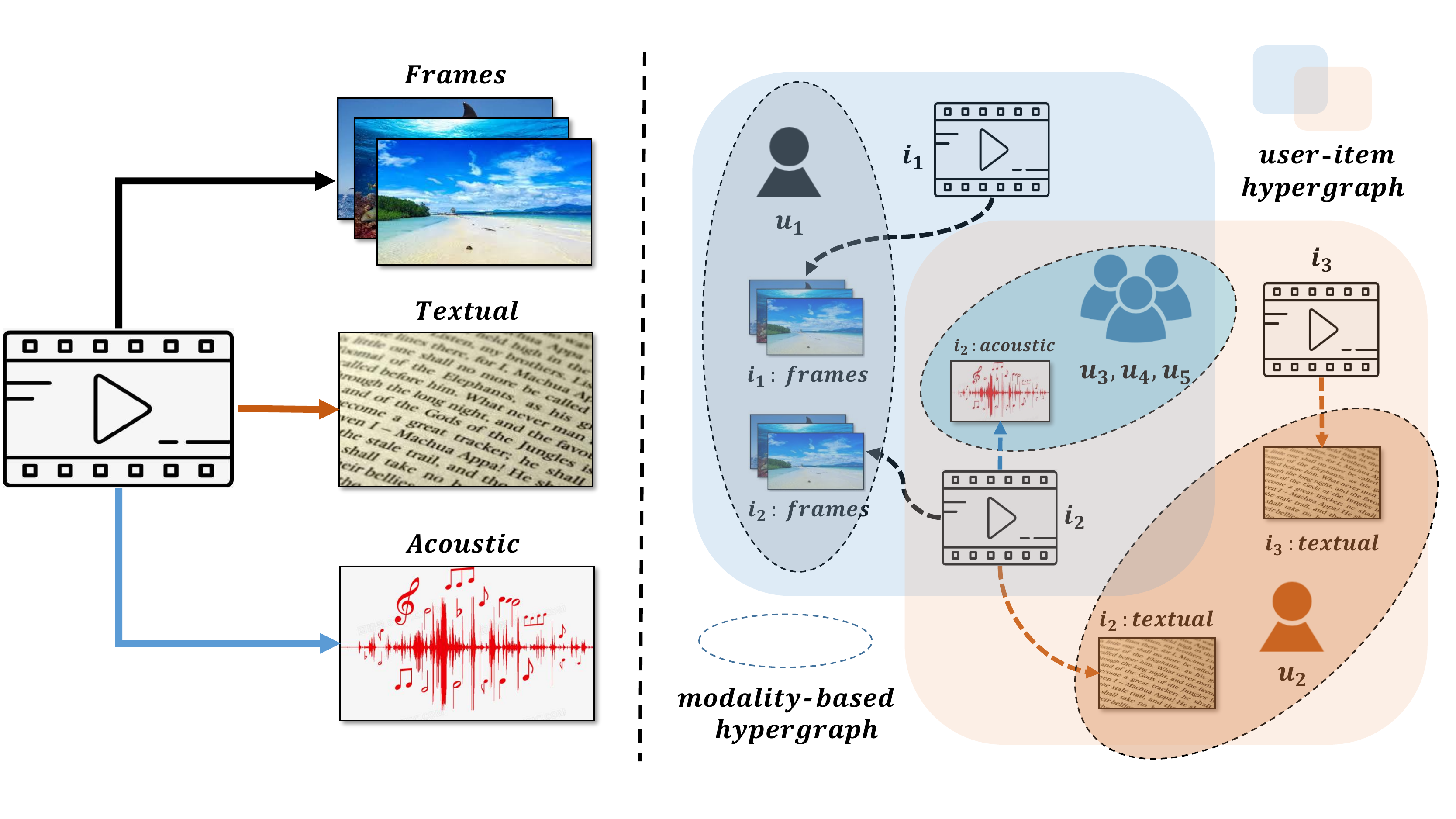}
  \vspace{-2em}
  \caption{An illustration of multi-modal user preferences.}
  \label{fig:toyexample_hyper}
  \vspace{-1.5em}
  \Description{}
\end{figure}

We rely on hypergraphs to address some of the challenges. Hypergraph \cite{bretto2013hypergraph,2021Multi} extends the concept of an edge in a graph and can connect more than two nodes. Inspired by the flexibility and expressiveness of hypergraphs, we use the concept to multi-modal feature modeling, and propose a new model based on modality-originated hypergraphs by which the sparsity issues between users and items under each modality can be alleviated. Figure \ref{fig:toyexample_hyper} is an example of the proposed modality-originated hypergraphs, where user \(u_1\) and user \(u_2\) both have interactions with multiple micro-videos, e.g., \(i_1\) and \(i_2\), in which each hyperedge can connect multiple item nodes on a single edge. Compared with a simple graph on which the degree of all edges is set to be 2, a hypergraph can encode high-order data correlation (beyond pairwise connections) using its degree-free hyperedges. Different from various modalities, we incorporate different multi-modal information, e.g., frames, acoustic, textual into user-item hypergraphs to help establish an in-depth understanding of user preferences. The reason for considering using hypergraphs in our work is due to the purpose of building modality-originated hypergraphs which can be treated as data argumentation technique.

We also construct hypergraphs considering both user and item. In Figure \ref{fig:toyexample_hyper}, user \(u_1\) cares more about frames of micro-video \(i_2\), whereas user \(u_2\) might be fond of the text content. Hence, different users might have different tastes on modalities of a micro-video. A group of users \(u_3, u_4\) and \(u_5\) click micro-video \(i_2\) due to the intriguing sound tracks. Such signals can be utilized to construct a group-aware hypergraph which is comprised of multiple users who share the same interest for the item. Inspired by the recent success of self-supervised learning (SSL)~\cite{Liu2020}, we utilize the mutual information maximization principle to learn the intrinsic data correlation~\cite{Zhou2020} to help construct the interests-based hypergraph where we represent a group of users with common preference on modal-specific content. Hence, in each modality (e.g., visual), we aggregate information from the group-aware hypergraph and incorporate them into user representations. According to group-aware hypergraph, each user has interactions with one of the item's modalities, while different items can be interacted with the same user. For example, user \(u_1\) likes \(i_1\)'s frames, and \(u_1\) will pay more attention to the visual-aspect of other items. Under such circumstances, we can also construct a homogeneous item-level hypergraph comprising of multiple items who have certain potential modality that appeal to the same user.

\begin{figure}
  \centering
  \includegraphics[width=\linewidth]{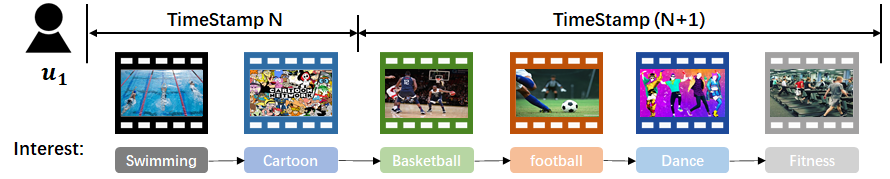}
  \vspace{-1em}
  \caption{Illustration of user \(u_1\)'s historical view records with micro-videos, which reflects the user's global view interests.}
  \vspace{-2em}
  \label{fig:toyexample}
\end{figure}
\hspace{-1em}

Generally, user preference evolves over time, and it is hence a sequential phenomenon. As shown in Figure~\ref{fig:toyexample}, user \(u_1\) has watched swimming and cartoon videos at timestamp \(N\), indicating that the user has two very different interests and we cannot capture the user's interests at the single time point. If at a new timestamp \(N+1\), basketball, football, dance and fitness videos have been selected by the same user. Then, we can infer that this user has more interests in sports than comedy. Under such circumstances, some researches consider users' interest as dynamic when designing CTR systems and have better users' interest models such as THACIL~\cite{2018Temporal}. Therefore, more user-behavior modeling methods are proposed for tacking this problem. There are RNN-based models~\cite{Li2019routing, hidasi2016sessionbased}, CNN-based models~\cite{2018Personalized}, transformer-based models~\cite{Pi2019} and memory network-based models~\cite{2018Collaborative,2021Where}.

To tackle the aforementioned problem, we propose HyperCTR, a novel temporal framework with user and item level hypergraphs to enhance CTR prediction. To explore the sequential correlations at different time slots, HyperCTR truncates the user interactions based on the timestamp to construct a series of hypergraphs. With a hypergraph convolutional network (HGCN), HyperCTR can aggregate the correlated users and items with direct or high-order connections to generate the dynamic embedding at each time slot. With change happening both over time and across users, the temporal and group-aware user embeddings are fed into a fusion layer to generate the final user representation. The prediction of an unseen interaction can be calculated as probability between the user and micro-video representations after MLP. We show the effectiveness of our framework on three publicly available datasets – Kuaishou, Micro-Video 1.7M (MV1.7M) and MovieLens. Our \textbf{key contributions} are: 1) We study the dynamics of user preference from two perspectives - time-aware and group-aware - and uncover the importance in exploiting the information interchange on various modalities to reflect user interests and affect CTR performance. 2) We propose a novel method HyperCTR framework with two types of modality-originated hypergraphs to generate users and items embeddings. Three of the unique aspects of the framework are a self-attention layer to capture the dynamic pattern in user-item bipartite interaction networks, a fusion layer to encode each interaction with both the temporal individual embeddings and group-level embeddings for final user pattern modeling and the CTR probability will be calculated by a MLP layer with the input of user- and item-level embeddings. 3) Extensive experiments on three public datasets demonstrate that our proposed model outperforms several state-of-the-art models. Due to anonymous requirements, the code link is invisible until paper acceptance.

\section{Our Novel HyperCTR Model}

\subsection{Preliminaries}
Our goal is learning user preferences from the hypergraph structure and predicting the probability that a user clicks the recommended entities. We denote \(U\) to represent the set of users and \(I\) represents the set of \(P\) items in an online platform. The item is characterised by various modalities, which are visual, acoustic, and textual. We also have historical interactions, such as ``click'' between users and items. We represent this interaction as a hypergraph \(\mathcal{G}(u,i)\), where \(u \in U\) and \(i \in I\) separately denote the user and item sets. A hyperedge, \(\mathcal{E}(u, i_1, i_2, i_3, ..., i_n)\) indicates an observed interaction between user \(u\) and multiple items \((i_1, i_2, i_3, ..., i_n)\) where hyperedge is assigned with a weight by \(\mathbf{W}\), a diagonal matrix of edge weights. We also have multi-modal information associated with each item, such as visual, acoustic and textual features. For instance, we denote \(M = \{v, a, x\}\) as the multi-modal tuple, where \(v\), \(a\), and \(x\) represent the visual, acoustic, and textual modalities, respectively.

Our hypothesis is that user preference also plays an important role. A user group \(y\) is associated with a user set \(C_y \in  U\) which can be used to represent a \(N\)-dimensional group-aware embedding. The member of groups might change over time. For each user \(u\), we denote the user's temporal behavior as \(B_u^c\) responding to the current time, and sequential view user behavior as \(B_u^s\) according to a time slot. We further utilize \(\mathcal{K}(B_u^c)\) and \(\mathcal{K}(B_u^s)\) to represent the set of items in the sequential behavior, respectively.

We explain some important terminologies below which includes temporal user-item interaction representation, group-aware hypergraph and item hypergraphs.

\begin{itemize}
    \item \textit{Definition 1 (Temporal User-item Interaction Representation)}
    
    Let a sequence \(\mathcal{S}(u, i_1, i_2, i_3, ...)\) indicate an observed interaction between user \(u\) and multiple items \((i_1, i_2, i_3, ...)\) occurring during a time slot \(t_n\). We denote \(\mathbf{E}_{I} = [\mathbf{e}_1,\mathbf{e}_2,...]\) as the set of items' static latent embeddings, which represent the set of items a user interacts with during this time slot. Each item in current sequence is associated with multi-modal features, which utilize \(M_{i_n}\) and it contains three-fold information about visual, acoustic and textual, denoted as \(v_{i_n}\), \(a_{i_n}\) and \(x_{i_n}\), respectively.

    \item \textit{Definition 2 (Group-aware Multi-Modal Hypergraph)}
    
    Let \(\mathcal{G}^{t_n}_{g}\) represent a hypergraph associated with \(i\)-th item at time slot \(t_n\). \(\mathcal{G}^{t_n}_{g} = \{V^{t_n}_{g}, \mathcal{E}^{t_n}_{g}, \mathbf{W}^{t_n}_{g}, \mathbf{H}^{t_n}_{g}\}\) is constructed based on the whole user-item interactions with multi-modal information. \(V^{t_n}_{g}\) represents the nodes of individual and the correlated item in \(\mathcal{G}^{t_n}_{g}\), \(\mathcal{E}^{t_n}_{g}\) denoted as a set of hyperedges. We are thus creating a link to users who have interactions with multiple modal list of items. Each \(\mathcal{G}^{t_n}_{g}\) is associated with an incidence matrix \(\mathbf{H}^{t_n}_{g}\) and it is also associated with a matrix \(\mathcal{W}^{t_n}_{g}\), which is a diagonal matrix representing the weight of the hyperedge \(\mathcal{E}^{t_n}_{g}\).
    
    \item \textit{Definition 3 (Item Homogeneous Hypergraph)}
    
    There are  three hyperedges in each \(\mathcal{G}^{t_n}_{g}\), which was defined in Definition 2. Let \(\mathcal{G}^{t_n}_{i}\)(\(\mathcal{G}^{t_n}_{i}\supseteq\{\mathbf{g}^{t_n}_{v},\mathbf{g}^{t_n}_{a},\mathbf{g}^{t_n}_{x}\}\)) represent a series of item homogeneous hypergraphs for each user group member. \(\mathcal{G}^{t_n}_{i} = \{V^{t_n}_{i}, \mathcal{E}^{t_n}_{i}, \mathbf{W}^{t_n}_{i}, \mathbf{H}^{t_n}_{i}\}\) is constructed based on each \(\mathcal{G}^{t_n}_{i}\) and describes a set of items that a user interacts with generated in the time slot \(t_n\). \(V^{t_n}_{i}\) represents the nodes of items and \(\mathcal{E}^{t_n}_{i}\) denotes a set of hyperedges, which is creating the link to items which have interactions with a user.
\end{itemize}

The group-aware hypergraph capture group member's preference, while item hypergraphs pay more attention to item-level high-order representation. Two types of hypergraphs are the fundamental for our temporal user-item interaction representation. We define our multi-modal hypergraph CTR problem as follow:
\begin{itemize}
    \item \textit{Problem 1} Click-Through Rate Prediction
    Given a target user intent sequence \(\mathcal{S}\), and its group-aware hypergraph \(\mathcal{G}^{t_n}_{g}\) and item hypergraph \(\mathcal{G}^{t_n}_{i}\), both of them depending on the time sequence \(T\), this problem can be formulated as a function \(f(u, \mathcal{G}^{t_n}_{g}, \mathcal{G}^{t_n}_{i}, i) \rightarrow y\) for a recommended item \(i\), where denotes \(y\) the probability that user clicks or not. 
\end{itemize}

\subsection{HYPERCTR Framework}
\begin{figure*}
\centering
\includegraphics[scale=0.6]{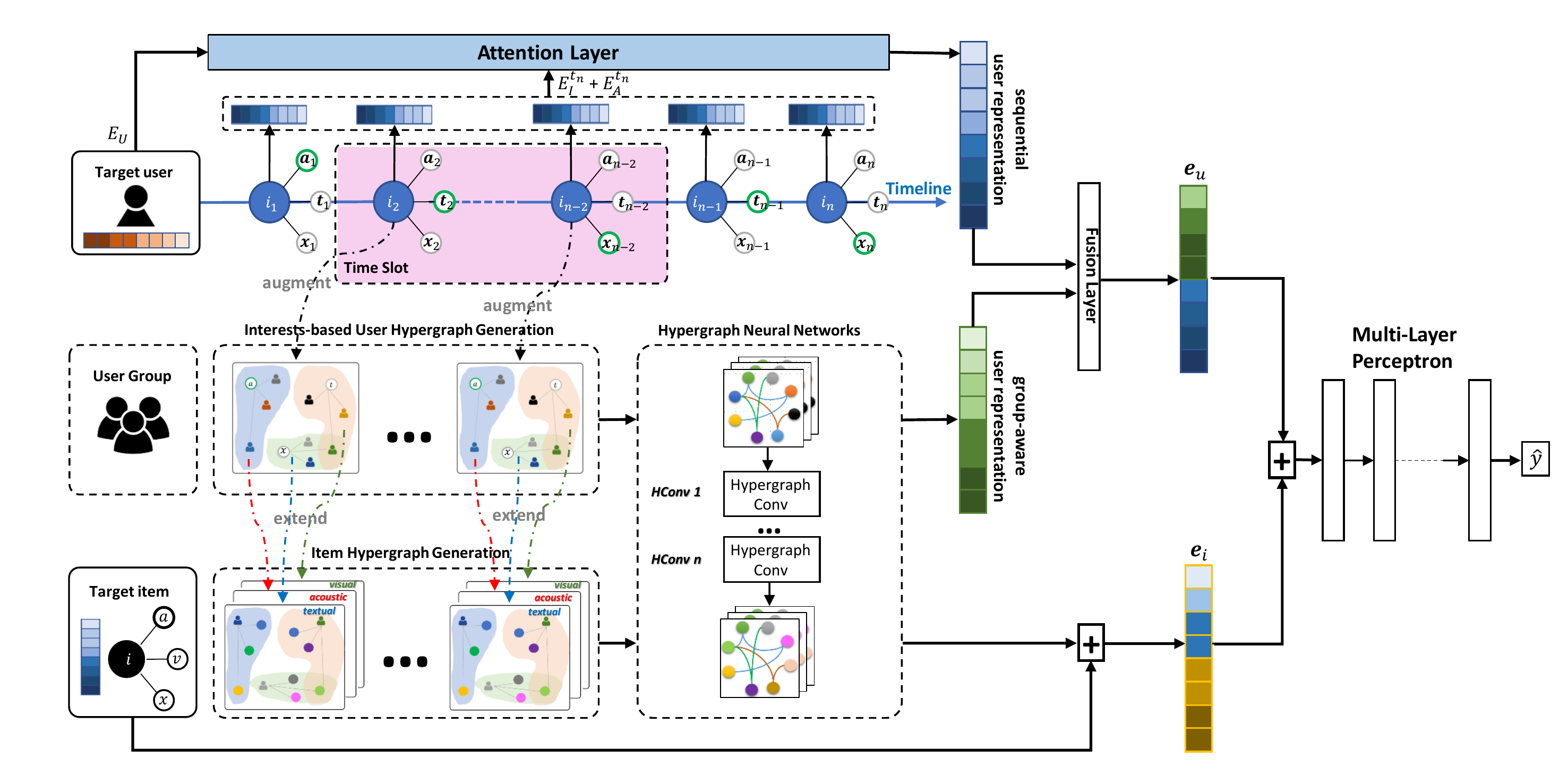}
\vspace{-2em}
\caption{The structure of HyperCTR: two views of hypergraphs are constructed based on user-item correlations at different time slot and the Hypergraph Neural Networks is able to capture the correlations in multi-hop connections. The attention layer can capture dynamic pattern in interaction sequences. Both the group-aware and sequential user embedding fuse to represent each individual, meanwhile, the target item embedding and a set of homogeneous item-item hypergraph embeddings are considered to learn the final prediction with the multi-layer perceptron.}
\label{fig:overview}
\vspace{-1em}
\end{figure*}

HyperCTR framework is illustrated in Figure~\ref{fig:overview}. The framework can be divided into four components: temporal user behavior attention module, interests-based user hyperedge generation module, item hypergraph construction module and prediction module. We illustrate the sequential user-item interactions in different timestamps from short-term and long-term granularity. The figure also shows that the target user has a pairwise relation with one item, while the item has multi-modal features such as visual, acoustic and textual. A user might have different tastes on modalities of an item, for example, a user is attracted by the frames, but might turn out to be disappointed with its poor sound tracks. Multiple modalities have varying contributions to user preferences. Each item can be treated as most current interactions from target user and the time-aware selection windows capture a time slot user behavior interacting on various items. All the short and long-term user intent and item embedding are fed into attention layer to represent each target user preference.

From group-level aspect, most item own more than one user. We extract item information from user-items sequential historical records and generate group-aware hyperedges. We can see in Figure~\ref{fig:overview} that there are three different colored areas. Every area denotes a hyperedge and a group of users connected by one unimodal feature in each hyperedge. We call this hyperedge Interest-based user hyperedge, and our task is to learn a user-interest matrix, leading to construct these hyperedges. Each hypergraph in the figure represents a group of users interacting with same item in the current time altogether and have different tendencies. We can then easily learn the group-aware information to enhance individual's representation. Besides, we have the opportunity to infer the preference of each user to make our prediction more accurate.

According to the group-level hyperedges, we can naturally find that each item can map to several users, while each user also has multiple interactions with various items. Here we cluster item information to build item hyperedges. There are several layers for each modality which extends from interests-based user hyperedges. The generation model will then go through the whole time period. We can now easily capture each higher-order structural relationship among items and enrich the representation of each items.

We leverage hypergraph convolutional operators to learn rich representation capturing local and higher-order structural relationships~\cite{Feng2019,li2021hyperbolic}. In the prediction module, we fuse group-aware user representation and sequential user representation. We then feed into a multi-layer perceptron and output the click-through rate prediction.

\vspace{-0.5em}

\subsubsection{Temporal User behavior Attention Module}
One user's historical interaction with items can span multiple times. A straightforward way is to apply RNN-type methods to analyze the sequence \(\mathcal{S}(u, i_1, i_2, i_3,\ldots)\). However, these models fail to capture both short-term and long-term dependencies. We thus perform a sequential analysis using the proposed temporal user behavior attention mechanism.

\textbf{Embedding Layer}
As depicted in Figure~\ref{fig:overview}, the long-term user interaction can be represented by all the items the user has interacted with in a certain time slot \(t_n\). In the user embedding mapping stage, to depict user behaviour features, we use their metadata and profiles and define an embedding matrix \(\mathbf{E}_U\) for each user \(\mathbf{u}_j\). We also maintain an item embedding matrix \(\mathbf{M}_{I} \in \mathbb{R}^{|\mathcal{I}|\times d}\) and a multi-modal attribute embedding matrix \(\mathbf{M}_{A} \in \mathbb{R}^{|\mathcal{A}|\times d}\). The two matrices project the high-dimensional one-hot representation of an item or multi-modal attribute to low-dimensional dense representations. Given a \(l\)-length time granularity sequence, we apply a time-aware slot window to form the input item embedding matrix \(\mathbf{E}_{I}^{t_n} \in \mathbb{R}^{l\times d}\). Besides, we also form an embedding matrix \(\mathbf{E}_{A}^{t_n} \in \mathbb{R}^{k\times d}\) for each item from the entire multi-modality attribute embedding matrix \(\mathbf{M}_{A}\), where \(k\) is the number of item modalities. The sequence representation \(\mathbf{E}^{t_n} \in \mathbb{R}^{n \times d}\) can be obtained by summing three embedding matrices: \(\mathbf{E}^{t_n}=\mathbf{E}_U + \mathbf{E}_{I}^{t_n} + \mathbf{E}_{A}^{t_n}\).

\textbf{Attention Layer}
We develop the sequential user behavior encoder by utilizing attention mechanism. We proposed to use self-attention layer, i.e., transformer which has also been applied in time series prediction~\cite{Rae2019}. In constrast to CNN, RNN-based approaches and Markov Chains-based models \cite{Kang2018}, we adopt self-attention as the basic model to capture the temporal pattern in user-items interaction sequence. A self-attention module generally consists of two sub-layers, i.e., a multi-head self-attention layer and a point-wise feed-forward network. The multi-head self-attention mechanism has been adopted for effectively extracting the information selectively from different representation subspaces~\cite{Zhou2020} and defined as:

\vspace{-0.8em}
\begin{equation}\begin{array}{l}
{ MultiHead }(Q, K, V)=\text { Concat }\left(\text { head }_{1}, \ldots, \text { head }_{\mathrm{h}}\right) W^{O}
\end{array}\end{equation}
\begin{equation}\begin{array}{l}
\text { head }_{\mathrm{i}}=\text { Attention }\left(Q W_{i}^{Q}, K W_{i}^{K}, V W_{i}^{V}\right)
\end{array}\end{equation}

where the projections are parameter matrices \(W_{i}^{Q} \in \mathbb{R}^{d \times d_k}\), \(W_{i}^{K} \in \mathbb{R}^{d \times d_k}\), \(W_{i}^{V} \in \mathbb{R}^{d \times d_v}\) and \(W^{O} \in \mathbb{R}^{h{d_v} \times d }\). The attention function is implemented by scaled dot-product operation:

\vspace{-1em}
\begin{equation}\begin{array}{l}
{ Attention }(Q, K, V)=\operatorname{softmax}\left(\frac{Q K^{T}}{\sqrt{d_{k}}}\right) V
\end{array}\end{equation}

where \((Q=K=V)=\mathbf{E}\) are the linear transformations of the input embedding matrix, and \(\frac{1}{\sqrt{d_{k}}}\) is the scale factor to avoid large values of the inner product, since the multi-head attention module is mainly build on the linear projections.

In addition to attention sub-layers, we applied a fully connected feed-forward network, denoted as \(\operatorname{FFN}(.)\), which contains two linear transformations with a ReLU activation in between.
\begin{equation}\begin{array}{l}
\operatorname{FFN}(x)=ReLU \left(0, x W_{1}+b_{1}\right) W_{2}+b_{2}
\end{array}\end{equation}
where \(W_1, b_1, W_2, b_2\) are trainable parameters.

\subsubsection{Hypergraph Convolution Network (HGCN)}
At each time slot, we aim to exploit the correlations among users and items for their high-order rich embeddings, in which the correlated users or items can be more complex than pairwise relationship, which is difficult to be modeled by a graph structure. On the other hand, the data representation tends to be multi-modal, such as the visual, text and social connections. To achieve that, each user should connect with multiple items with various modality attributes, while each item should correlated with several users. This naturally fits the assumption of the hypergraph structure for data modeling. Compared with simple graph, on which the degree for all edges is mandatory to be 2, a hypergraph can encode high-order data correlation using its degree-free hyperedges \cite{Feng2019}. In our work, we construct a \(\mathcal{G}(u,i)\) to present user-item interactions over different time slots. Then, we aim to distill some hyperedges to build user interest-based hypergraph \(\mathcal{G}_{g}^{t_n}\)and item hypergraph \(\mathcal{G}_{i}^{t_n}\) to aggregate high-order information from all neighborhood. We concatenate the hyperedge groups to generate the hypergraph adjacent matrix \(\mathbf{H}\). The hypergraph adjacent matrix \(\mathbf{H}\) and the node feature are fed into a convolutional neural network (CNN) to get the node output representations. We build a hyperedge convolutional layer \(f(\mathbf{X}, \mathbf{W}, \mathbf{\Theta})\) as follows:
\begin{equation}\begin{array}{l}
\mathbf{X}^{(l+1)}=\sigma\left(\mathbf{D}_{v}^{-1 / 2} \mathbf{H} \mathbf{W} \mathbf{D}_{e}^{-1} \mathbf{H}^{\top} \mathbf{D}_{v}^{-1 / 2} \mathbf{X}^{(l)} \Theta^{(l)}\right)
\end{array}\end{equation}
where define \(\mathbf{X}, \mathbf{D}_{v}, \mathbf{D}_{e}\) and \(\Theta\) is the signal of hypergraph at \(l\) layer, \(\sigma\) denotes the nonlinear activation function. The GNN model is based on the spectral convolution on the hypergraph.

\subsubsection{Prediction Module and Losses}
We want to incorporate both user sequential embeddings and group-aware high-order information for a more expressive representation of each user in the sequence. We propose the fusion layer to generate the representation of user \(u\) at \(t_n\). Existing works on multiple embeddings use concatenation as fusion~\cite{li2019spam}, resulting in suboptimal interactions. We utilize the fusion process that transforms the input representations into a heterogeneous tensor \cite{mai2020modality,2017SPTF,2020Heterogeneous}. We use the user sequential embedding \(\mathbf{E}^{t_n}\) and group-aware hypergraph embedding \(\mathbf{E}^{t_n}_{g}\). Each vector \(\mathbf{E}\) is augmented with an additional feature of constant value equal to 1, denoted as \(\mathbf{E}=(\mathbf{E},1)^{T}\). The augmented matrix \(\mathbf{E}\) is projected into a multi-dimensional latent vector space by a parameter matrix \(\mathbf{W}\), denoted as \(\mathbf{W}^T \mathbf{E}_m\).  Therefore, each possible multiple feature interaction between user and group-level is computed via outer product, \(\mathcal{T}=f\left(\mathbf{W}^{T} \cdot \tilde{\mathbf{E}}_{m}\right)\), expressed as:

\begin{equation}\begin{array}{l}
\mathcal{T}_{U}=\mathbf{W}^{T} \cdot\left(\mathbf{E}^{t_n} \otimes \mathbf{E}_{g}^{t_n}\right)
\end{array}\end{equation}
Here $\otimes$ denotes outer product, \(\tilde{\mathbf{E}_{m}}\) is the input representation from user and group level. It is a two-fold heterogeneous user-aspect tensor \(\mathcal{T}_{U}\), modeling all possible interrelation, i.e., user-item sequential outcome embeddings \(\mathbf{E}^{t_n}\) and group-aware aggregation features \(\mathbf{E}^{t_n}_{g}\).

When predicting the CTR of user for items, we take both sequential user embedding and item embedding into consideration. We calculate the user-level probability score \(y\) to a candidate item \(i\), to clearly show how the function \(f\) works. The final estimation for the user CTR prediction probability is calculated as:
\begin{equation}\begin{array}{l}
\hat{y}=f(\boldsymbol{e}_{u}, \boldsymbol{e}_{i};\boldsymbol{\Theta})
\end{array}\end{equation}
where \(\boldsymbol{e}_{u}\) and \(\boldsymbol{e}_{i}\) denote user and item-level embeddings, respectively. \(f\) is the learned function with parameter \(\boldsymbol{\Theta}\) and implemented as a multi-layer deep network with three layers, whose widths are denoted as \(\{D_1,D_2,\ldots,D_N\}\) respectively. The first and second layer use \(ReLU\) as activation function while the last layer uses sigmoid function as \(\operatorname{Sigmoid}(x)=\frac{1}{1+e^{-x}}\). As for the loss function, we take an widely used end-to-end training approach, Cross Entropy Loss\cite{2018Bidding,zhou2018deep,Feng2019deep}, and it is formulated as:
\begin{equation}
L(\boldsymbol{e}_{u}, \boldsymbol{e}_{i})=y \log \sigma(f(\boldsymbol{e}_{u}, \boldsymbol{e}_{i}))+(1-y) \log (1-\sigma(f(\boldsymbol{e}_{u}, \boldsymbol{e}_{i})))
\label{eq:loss}
\end{equation}
where \(y \in \{0, 1\}\) is ground-truth that indicates whether the user clicks the micro-video or not, and \(f\) represents the multi-layer deep network.

\subsection{Hypergraph Generation Modules}
We aim to distill the user-level hypergraph group to enhance the representations of input data. We adopt a pre-training way to learn user group latent preference correlation to different modalities from items. However, as model trained is prone to suffer from unlabelled data problem, there is no explicit information to associate user and each item's modality. We further incorporate additional self-supervised signals with mutual information to learn the intrinsic data correlation~\cite{Zhou2020,Liu2020}.

\subsubsection{Interest-based User Hypergraph Generation Modeling}

We aim to utilize self-supervised learning for the user-interest matrix \(\mathbf{F} \in \mathbb{R}^{L \times d}\), where \(L\) denote the user counts and \(d\) denote the number of multi-modalities according to items. We trained the weights \(\{\theta_{a}, \theta_{b}, \theta_{c}\}\) for each modalities. We define \(\{\alpha,\beta,\gamma\}\) to denote the degree of interest of each modalities from the item features. A threshold \(\delta\) was applied to measure which modality contributes the most for user-item interaction. We first maximize the mutual information between users \(u\) and item's multi-modal attributes \(M_{i_n}^{t_n}\). For each user and item, the metadata and attributes provide fine-grained information about them. We aim to fuse user and multimodal-level information through modeling user-multimodal correlation. It is thus expected to inject useful multi-modal information into user group representation.
Given an item \(i\) and the multi-modal attributes embedding matrix \(\mathbf{M}_{i_i}^{t_n} \in \mathbb{R}^{|\mathcal{A}|\times d}\), we treat user, item and its associated attributes as three different views denoted as \(\mathbf{E}_{U}\), \(\mathbf{E}_{I}^{t_n}\) and \(\mathbf{E}_{A}^{t_n}\). Each \(\mathbf{E}_{A}^{t_n}\) is associated with a embedding matrix \(M_k \in M_{i_n}^{t_n}=\{v_{i_n}^{t_n},a_{i_n}^{t_n},x_{i_n}^{t_n}\}\). We design a loss function by the contrastive learning framework that maximizes the mutual information between the three views. Following Eq~\ref{eq:loss}, we minimize the User Interest Prediction (UIP) loss by:

\vspace{-0.5em}
\begin{small} 
\begin{equation}\begin{array}{l}
L_{U I P}\left(u, i, \mathbf{E}_{A_i}\right)=\mathbb{E}_{a_{j} \in \mathbf{E}_{A_i}}\left[f\left(u, i, a_{j}\right)-\log \sum_{\tilde{a} \in \mathbf{E}_{A} \backslash \mathbf{E}_{A_i}} \exp (f(u, i, \tilde{a}))\right]
\end{array}\end{equation}
\end{small}
where we sample negative attributes \(\tilde{a}\) that enhance the association among user, item and the ground-truth multi-modal attributes, "\(\backslash\)" defines set subtraction operation. The function \(f(\cdot,\cdot,\cdot)\) is implemented with a simple bilinear network:
\begin{equation}\begin{array}{l}
f\left(u, i, a_{j}\right)=\sigma\left[\left(\mathbf{E}_{I}^{\top} \cdot \mathbf{W}_{U I P} \cdot \mathbf{E}_{A_{j}}\right) \cdot \mathbf{E}_{U} \right]
\end{array}\end{equation}
where \(\mathbf{W}_{U I P} \in \mathbb{R}^{d \times d}\) is a parameter matrix to learn and \(\sigma(.)\) is the sigmoid function. We define the loss function \(L_{U I P}\) for a single user, which will can be extended over the user set in a straightforward way.
The outcome from \(f(.)\) for each user can be constructed as a user-interest matrix \(\mathbf{F}\) and compared with the threshold \(\delta\) to output the \(L\)-dimensions vector \(\mathbf{v} \in \mathbb{R}^{1 \times L}\).

\subsubsection{Item Hypergraph Construction}
We exploit how to transform a sequential user-item interactions into a set of homogeneous item-level hypergraph. We construct a set of homogeneous hypergraphs \(\mathcal{G}_I\), from node sets \(I\) as follow:
\begin{equation}\begin{array}{l}
\mathcal{G}_I = \{\mathcal{G}_{I,\text{group}}, \mathcal{G}_{I,1},\ldots,\mathcal{G}_{I,Q}\}
\end{array}\end{equation}
where \(\mathcal{G}_{I,j} = \{I, \mathcal{E}_{I,j}\}\), and \(\mathcal{E}_{I,j}\) denote hyperedges in \(\mathcal{G}_{I,j}\). Note that all the homogeneous hypergraphs in \(\mathcal{G}_I\) share the same node set \(I\). For a node \(i \in I\), a hyperedge introduced in \(\mathcal{E}_{I,j}\) of \(\mathcal{G}_{I,j}\), which connects to \(\{i | i \in I, (u, i) \in \mathcal{E}_{T_n}\}\), i.e., the vertices in \(I\) that are directly connected to \(u\) by \(\mathcal{E}_{T_n}\) in time period \(T_n\). According to Figure~\ref{fig:overview}, in the user-item sequential interaction network, the user \(u\) clicks three items \(v\), which corresponds to a hyperedge that connects these three items in the homogeneous hypergraph \(\mathcal{G}_I\). The special homogeneous hypergraph \(\mathcal{G}_{I,\text{group}} \in \mathcal{G}_I\) are defined as \(G\left(I, \bigcup_{j=1}^{k} \mathcal{E}_{I, j}\right)\). Note that the cardinalities of hyperedge sets in the constructed hypergraph are: \(|\mathcal{E}_{I, j}| \leq |U|\) and \(|\mathcal{E}_{I, \text{group}}| \leq k|U|\) for \(j \leq k\). The total number of hyperedges in the homo-hypergraph is proportional to the number of nodes and edge types in the input sequence: \(O(k(|I|+|V|))\). Thus, the transformation easily scales to large inputs.

\subsubsection{Information Augmentation}
The increasing data sparsity problem is one of our main motivations in tackling with CTR prediction task. To address the interaction sparsity problem, some information augmentation methods have been proposed~\cite{2018Subgraph, 2020Gemini}, however, they only consider in the case of single modality and cannot handle the scenarios with multi-modal features. We propose two data augmentation strategies, which use user behavior information and item multi-modal information to learn the subgraph embedding. We transform the initial user-item heterogeneous hypergraph into two homogeneous hypergraphs from the perspective of users and items respectively. 

\textbf{User Behavior Information Augment Strategy}
We have utilized temporal user interaction logs to represent user-level embedding. However, the heterogeneous nature between users and items aggravates the difficulty in network information fusion. A common observation is that the user usually interacts with only a small number of items while an item can only be exposed to a small number of users, which results in a sparse user-item network and limits the effectiveness of embedding representation. To mitigate the issue, we utilize the self-supervised user interest matrix \(\mathbf{F}\) to build the user-user homogeneous graphs, which contains multiple hyperedges, and is regarded as hypergraph. It is denoted as \(\mathcal{G}^{t_n}_{g}\) mentioned in Definition 2.

\textbf{Item Multi-modal Information Augment Strategy}
It is a common observation that if two users both link to the same modality of items, then they have some common interest~\cite{2016Zhang}. We are thus motivated to add an edge between them in \(\mathcal{G}^{t_n}_{g}\). Similarly, if some items link to the same set of users, they share the same target user group. We thus add an hyperedge between them in \(\mathcal{G}^{t_n}_{i}\).

According to the two information augmentation strategies, we transform the first-order neighbor relations of user-item to second-order neighbor relations of user-user and item-item, and represent the complex relationship as a multiple hypergraph structure. Compared with single hop neighbors, in our case nodes have more hop neighbors, which can be used to alleviate the problem of graph sparsity. The items in each hyperedge in \(\mathcal{G}^{t_n}_{i}\) maintain some intrinsic attribute correlation due to which they connect with the same user preference. Adding edge information while aggregating information from neighbor nodes can exchange heterogeneous topology information between \(\mathcal{G}^{t_n}_{g}\) and \(\mathcal{G}^{t_n}_{i}\). The information fusion processes on the two graphs are interdependent.

\section{Experiments and Results}
\subsection{Experimental Settings}
\subsubsection{Datasets}
Existing CTR prediction models mostly utilize unimodal datasets \cite{Liu2019,Song2020,Liu2020,Qin2020}. In contrast, we introduce multiple modalities into CTR prediction. As mentioned above, micro-video datasets contain rich multimedia information and include multiple modalities - visual, acoustic and textual. We experimented with three publicly available datasets which are summarized in Table~\ref{tab:dataset}.
\begin{table}
\centering
\caption{Statistics of the dataset. (v, a and t denote the dimensions of visual, acoustic, and textual modalities, respectively.)}
\vspace{-1em}
\label{tab:dataset}
\resizebox{8.5cm}{!}{ 
\begin{tabular}{c|c|c|c|c|c|c|c} \toprule
\hline
Dataset & \#Items & \#Users & \#Interactions & Sparsity & v. & a. & t. \\ \hline \hline
Kuaishou & 3,239,534 & 10,000 & 13,661,383 & 99.98\% & 2048 & - & 128 \\
MV1.7M  & 1,704,880 & 10,986 & 12,737,619 & - & 128 & 128 & 128   \\
Movielens & 10,681 & 71,567  & 10,000,054 & 99.63\% & 2048 & 128 & 100 \\ \hline
\bottomrule
\end{tabular}}
\vspace{-1em}
\end{table}

\textbf{Kuaishou}: This dataset is released by the Kuaishou \cite{Li2019routing}. There are multiple interactions between users and micro-videos. Each behaviour is also associated with a timestamp, which records when the event happens.

\textbf{Micro-Video 1.7M}: This dataset was proposed in \cite{2018Temporal}. The interaction types include ``click'' and ``unclick''. Each micro-video is represented by a 128-dimensional visual embedding vector of its thumbnail. Each user's historical interactions are sorted in chronological order.

\textbf{MovieLens}: The MovieLens dataset is obtained from the MovieLens 10M Data\footnote{http://files.grouplens.org/datasets/movielens/}. We assume that a user has an interaction with a movie if the user gives it a rating of 4 or 5. We use ResNet\cite{he2015deep}, VGGish \cite{hershey2017cnn} and Sentence2Vector \cite{logeswaran2018efficient} to handle the visual features, acoustic modality and textual information respectively.

\subsubsection{Baseline Models}
We compare our model with strong baselines from both sequential CTR prediction and recommendation. Our comparative methods are: 1) \textbf{GRU4Rec} \cite{hidasi2016sessionbased} based on RNN. 2) \textbf{THACIL} \cite{2018Temporal} is a personalized micro-video recommendation method for modeling user's historical behaviors. 3) \textbf{DSTN} \cite{2019DeepSpatio} learns the interactions between each type of auxiliary data and the target ad, to emphasize more important hidden information, and fuses heterogeneous data in a unified framework. 4) \textbf{MIMN} \cite{Pi2019} is a novel memory-based multi-channel user interest memory network to capture user interests from long sequential behavior data. 5) \textbf{ALPINE} \cite{Li2019routing} is a personalized micro-video recommendation method which learns the diverse and dynamic interest, multi-level interest, and true negative samples. 6) \textbf{AutoFIS} \cite{Liu2020AutoFIS} automatically selects important \(2^{nd}\) and \(3^{rd}\) order feature interactions. The proposed methods are generally applicable to many factorization models and the selected important interactions can be transferred to other deep learning models for CTR prediction. 7) \textbf{UBR4CTR} \cite{Qin2020} has a retrieval module and it generates a query to search from the whole user behaviors archive to retrieve the most useful behavioral data for prediction.

\subsubsection{Parameter Settings}
We randomly split all datasets into training, validation, and testing sets with 7:2:1 ratio, and create the training triples based on random negative sampling. For testing set, we pair each observed user-item pair with 1000 unobserved micro-videos that the user has not interacted with before. 

For our baseline methods, we use the implementation and settings provided in their respective papers. More details show as follow items and Table~\ref{tab:baseline}.

\begin{itemize}
    \item \textbf{GRU4Rec} We applies GRU to model user click sequence for reproduce this model. We represent the items using embedding vectors rather than one-hot vectors.
    \item \textbf{THACIL} The number of micro-videos per user is set to 160. The temporal block size is set to 20. For users having more items than 160, we just preserve as much as 160 items. For users having less items, we pad all-zero vectors to augment.
    \item \textbf{DSTN} We set the dimension of the embedding vectors for each feature as 10, because the number of distinct features is huge. We set the number of fully connected layers in DSTN is 2, each with dimensions 512 and 256.
    \item \textbf{MIMN} Layers of FCN (fully connected network) are set by 200 × 80 × 2. The number of embedding dimension is set to be 16. The number of hidden dimension for GRU in MIU is set to be 32. We take AUC as the metric for measurement of model performance.
    \item \textbf{ALPINE} We utilized the 64-d visual embedding to represent the micro-video. The length of users’ historical sequence is set to 300. If it exceeds 300, we truncated it to 300; otherwise, we padded it to 300 and masked the padding in the network.
    \item \textbf{AutoFIS} We implement the two-stage algorithm AutoFIS to automatically select important low-order and high-order feature interactions with FM-based model.
    \item \textbf{UBR4CTR} The datasets are processed into the format of comma separated features. A line containing user, item and context features is treated as a behavior document.
\end{itemize}

\begin{table}
\centering
\caption{Parameter Settings}
\vspace{-1em}
\label{tab:baseline}
\begin{tabular}{c|c|c|c} \toprule
\hline
Methods & \#Batch size & \#Dropout & \#Learning rate  \\ \hline \hline
GRU4Rec  & 200 & 0.1 & 0.05 \\
THACIL  & 128 & 0.2 & 0.001 \\
DSTN  & 128 & 0.5 & 0.001\\ 
MIMN  & 200 & 0.2 & 0.001\\ 
ALPINE  & 2048 & 0.3 & 0.001 \\ 
AutoFIS  & 2000 &  0.6 & 0.005\\ 
UBR4CTR  & 200 & 0.5 & 0.001 \\ \hline
\end{tabular}
\end{table}

In HyperCTR and all its variants use Adam optimizer. For training, we randomly initialize model parameters with a Gaussian distribution and use the ReLU as the activation function. We then optimized the model with stochastic gradient descent (SGD). We search the batch size in {128, 256, 512}, the the latent feature dimension in {32, 64, 128}, the learning rate in {0.0001, 0.0005, 0.001.0.005, 0.01} and the regularizer in {0, 0.00001, 0.0001, 0.001, 0.01, 0.1}. As the findings are consistent across the dimensions of latent vectors, we have shown the result of 64, a relatively large number that returns good performance whose details can be found sensitivity analysis.

\subsubsection{Evaluation Metrics}
We evaluate the CTR prediction performance using two widely used metrics. The first one is Area Under ROC curve (AUC) which reflects the pairwise ranking performance between click and non-click samples. The other metric is log loss. Log loss is to measure the overall likelihood of the test data and has been widely used for the classification tasks \cite{Ren2018, Ren2016}.

\begin{table}
\caption{The overall performance of different models on Kuaishou, Micro-Video 1.7M and MovieLens datasets in \%.}
\vspace{-1em}
\centering
\footnotesize
\begin{tabular}{|l|cc|cc|cc|}
\hline
\multirow{2}{*}{Method} & \multicolumn{2}{c|}{Kuaishou} & \multicolumn{2}{c|}{MV1.7M} & \multicolumn{2}{c|}{MovieLens} \\& AUC  & Logloss  & AUC  & Logloss  & AUC & Logloss \\ \hline\hline
GRU4Rec & 0.7367 & 0.5852  & 0.7522& 0.6613   & 0.7486 & 0.6991 \\
THACIL  & 0.6640 & 0.5793   & 0.6842 & 0.6572   & 0.6720 & 0.6791  \\
DSTN & 0.7722 &  0.5672  & 0.7956 & 0.6492    & 0.8008 & 0.6162    \\
MIMN  & 0.7593 & 0.5912    & 0.7486&  0.6862   &  0.7522 &     0.6751\\ 
ALPINE  &  0.6840 &  \uline{0.5632}  & 0.7130 & 0.6591   &0.7390 & 0.6163 \\ 
AutoFIS  & \uline{0.7870} & 0.5756 & 0.8010 & \uline{0.5404} & 0.7983 & \uline{0.5436} \\
UBR4CTR  & 0.7520 &0.5710 & \uline{0.8070} & 0.5605 & \uline{0.8050} & 0.5663 \\ \hline
\textbf{HYPERCTR}  & \textbf{0.8120} & \textbf{0.5548} & \textbf{0.8670} & \textbf{0.5160} & \textbf{0.8360} & \textbf{0.5380}  \\ \hline
Improv.(\%) & 3.18\%  &1.49\% &7.43\% &4.51\% &3.85\% &1.03\%  \\ \hline
\end{tabular}
\vspace{-1em}
\label{tab:overall}
\end{table}

\subsection{Quantitative Performance Comparison} 
Table~\ref{tab:overall} presents the AUC score and Logloss values for all models. When different modalities re used, all models show an improved performance when the same set of modalities containing visual, acoustic and textual features are used in MV1.7M and MoiveLens(10M). We also note that: (a) the performance of our model has improved significantly compared to the best performing baselines. AUC is improved by 3.18\%, 7.43\% and 3.85\% on three datasets, respectively, and Logloss is improved by 1.49\%, 4.51\% and 1.03\%, respectively; and (b) the improvement in our model demonstrates that the unimodal features do not embed enough temporal information which the baselines cannot exploit effectively. The baseline methods cannot perform well if the patterns that they try to capture do not contain multi-modal features in the user-item interaction sequence.

\subsection{HyperCTR Component Analysis}
\subsubsection{Role of Multimodality }
To explore the effect of different modalities, we compare the results on different modalities on the three datasets, as shown in Table~\ref{tab:ablation}. We make the following observations: 1) Our main method outperforms those with single-modal features on three datasets. It demonstrates that representing users with multi-modal information achieves a better performance. It also demonstrates that the construction of hyperedges can capture user's modal-specific preference from graph information. 2) The visual-modal is the most effective one among three modalities. It can be naturally understood that if a user clicks what to watch, one usually pays more attention to the visual information than other modality. 3) The acoustic-modal shows more important information for user click compared with the textual features. This is expected as the background music is more attractive to users. 4) Textual modality contributes least towards click-through rate prediction. However, in MovieLens data corpus, this modality has smaller gap with the other modalities. This is because the text in MovieLens is highly related to the content. 4) Compared with GCN, our proposed model achieved better performance in all datasets. As shown in Table~\ref{tab:ablation}, based on Kuaishou datasets, when only two features are used for hypergraph, our model can still obtain slight improvement. With more features in the other two datasets, our model achieves much better performance compared with GCN. This phenomenon is consistent with our argument that when multi-modal features are available, hypergraph has the advantage of combining such multi-modal information in the same structure by its flexible hyperedges.

\begin{table}
\caption{Performance in terms of AUC \& Logloss w.r.t different modalities on the three datasets in \%.}
\vspace{-1em}
\centering
\footnotesize
\begin{tabular}{|l|cc|cc|cc|}
\hline
\multirow{2}{*}{Method} & \multicolumn{2}{c|}{Kuaishou} & \multicolumn{2}{c|}{MV1.7M} & \multicolumn{2}{c|}{MovieLens} \\& AUC  & Logloss  & AUC  & Logloss  & AUC & Logloss \\ \hline\hline
multi-modal     & \textbf{0.8120} & \textbf{0.5548} & \textbf{0.8670} & \textbf{0.5160} & \textbf{0.8360} & \textbf{0.5380} \\
visual-modal    & 0.8110 & 0.5560 & 0.8567 & 0.5167 & 0.8259 & 0.5376       \\
acoustic-modal  &   -    &    -   & 0.8260 & 0.5171 & 0.8134 & 0.5373       \\
textual-modal   & 0.7720 & 0.5756 & 0.8158 & 0.5175 & 0.8123 & 0.5364       \\ 
(-) hypergraph &  0.8034 & 0.5554 & 0.8137 & 0.5426 & 0.8064 & 0.5673 \\\hline
\end{tabular}
\label{tab:ablation}
\vspace{-1em}
\end{table}

\subsubsection{Role of HGCN Layers}
\begin{figure}
  \centering
  \includegraphics[width=\linewidth]{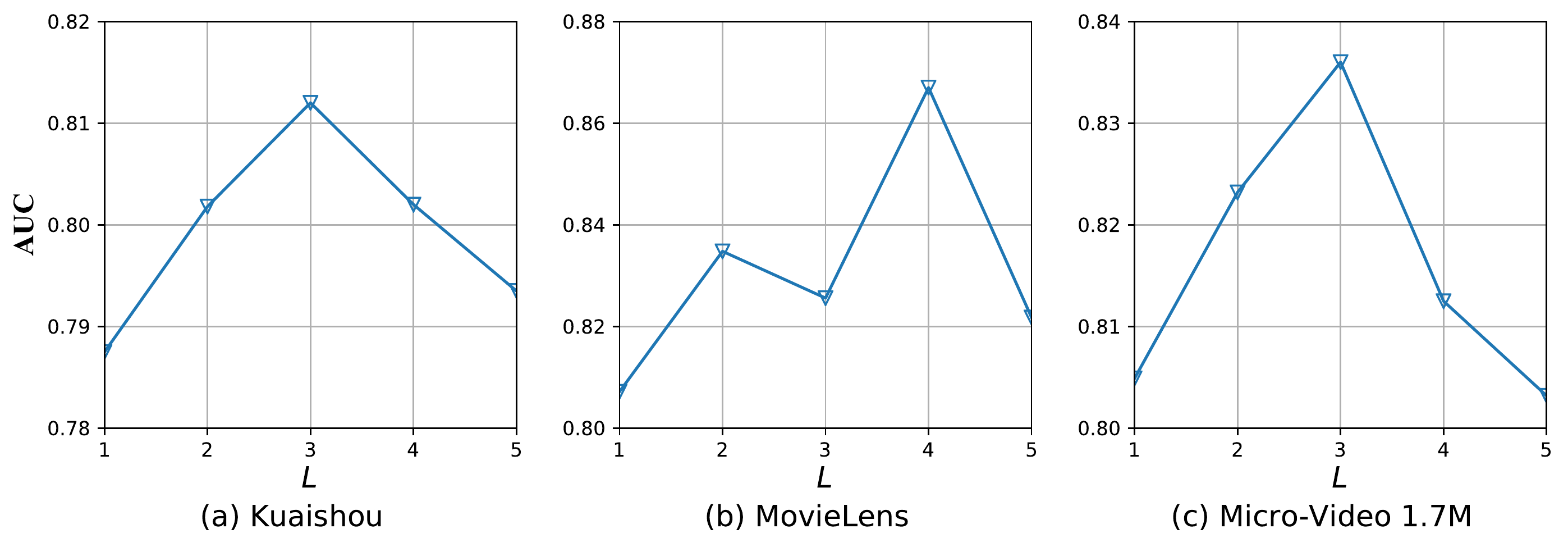}
  \caption{Performance comparison with different number of HGCN layers under AUC}
  \vspace{-1em}
  \label{fig:layers}
\end{figure}

To explore how the high-order connections in the hypergraph can help to uncover hidden item correlations and thus contribute to the final prediction. We compare the performance of HyperCTR by varying the number of hypergraph convolutional layers. As shown in Figure~\ref{fig:layers}, when we apply only one convolution layer for our sequential model, each node embedding aggregates only information from others connected with them directly by the hyperedge. Our model performs poorly in all three datasets. By stacking three HGCN layers, it can bring in significant improvement compared with a model with just one convolution layer. We can infer that HGCN are useful options for extracting expressive item semantics and it is important to take the high-order neighboring information in hypergraph into consideration. On Kuaishou and MV1.7M, since the data is very sparse, it is not necessary to further increase the number of convolutional layers. Three HGCN layers are enough for extracting the user- and item-level semantics at different time slots. On MovieLens, more convolutional layers can further improve the embedding process. This demonstrates the effectiveness of hypergraph and HGCN in modeling the temporal user and item correlations.

\vspace{-0.5em}

\subsubsection{Role of Time Granularity}
\begin{figure}
  \centering
  \includegraphics[width=\linewidth]{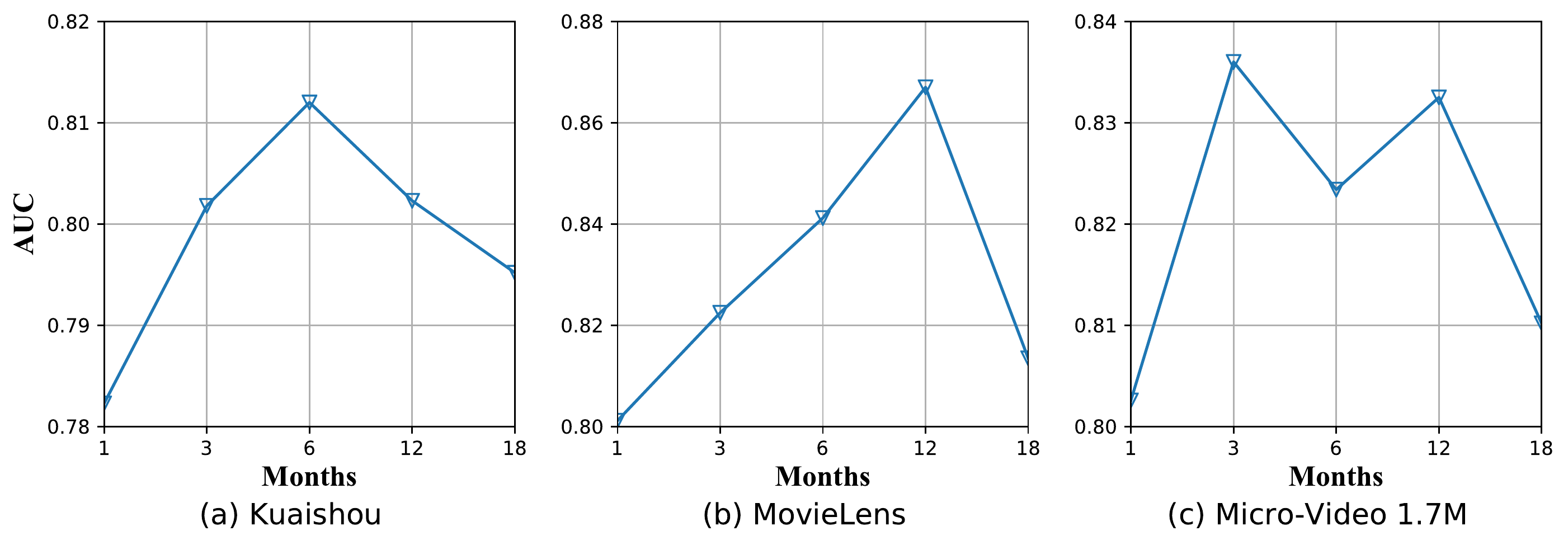}
  \vspace{-2em}
  \caption{Performance comparison with various time granularity under AUC}
  \label{fig:time}
  \vspace{-1em}
\end{figure}

An important parameter which can effect the performance of HyperCTR is the granularity of the time slot. According to Figure~\ref{fig:time}, we show the performance of the proposed model by varying the granularity from 1 month to 18 months. When the granularity is small, we find that the model cannot achieve the best performance since the interactions are extremely sparse and not sufficient for building up a set of expressive user and item embeddings. While enlarging the granularity, we find that the performance of HyperCTR is increasing in all the datasets. In Kuaishou datasets, it reaches the best performance when the time granularity is set to half a year. However, for MovieLens, the optimized granularity is almost one year since the item in MovieLens is movie, it propagation speed is relatively slow and the impact time is relatively long. In MV1.7M datasets, the optimized granularity is around three months, which is smaller than that for the other datasets since the micro-video sharing platform attracts more interactions for each time slot for the temporal user preference representations. If we further enlarge the granularity, the performance will decrease since it underestimates the change of user preference and may introduce noise to the model.

\subsection{HyperCTR Model Parameter Study}
\subsubsection{Hyperparametr Sensitivity Analysis}
\begin{figure}
  \centering
  \includegraphics[width=\linewidth]{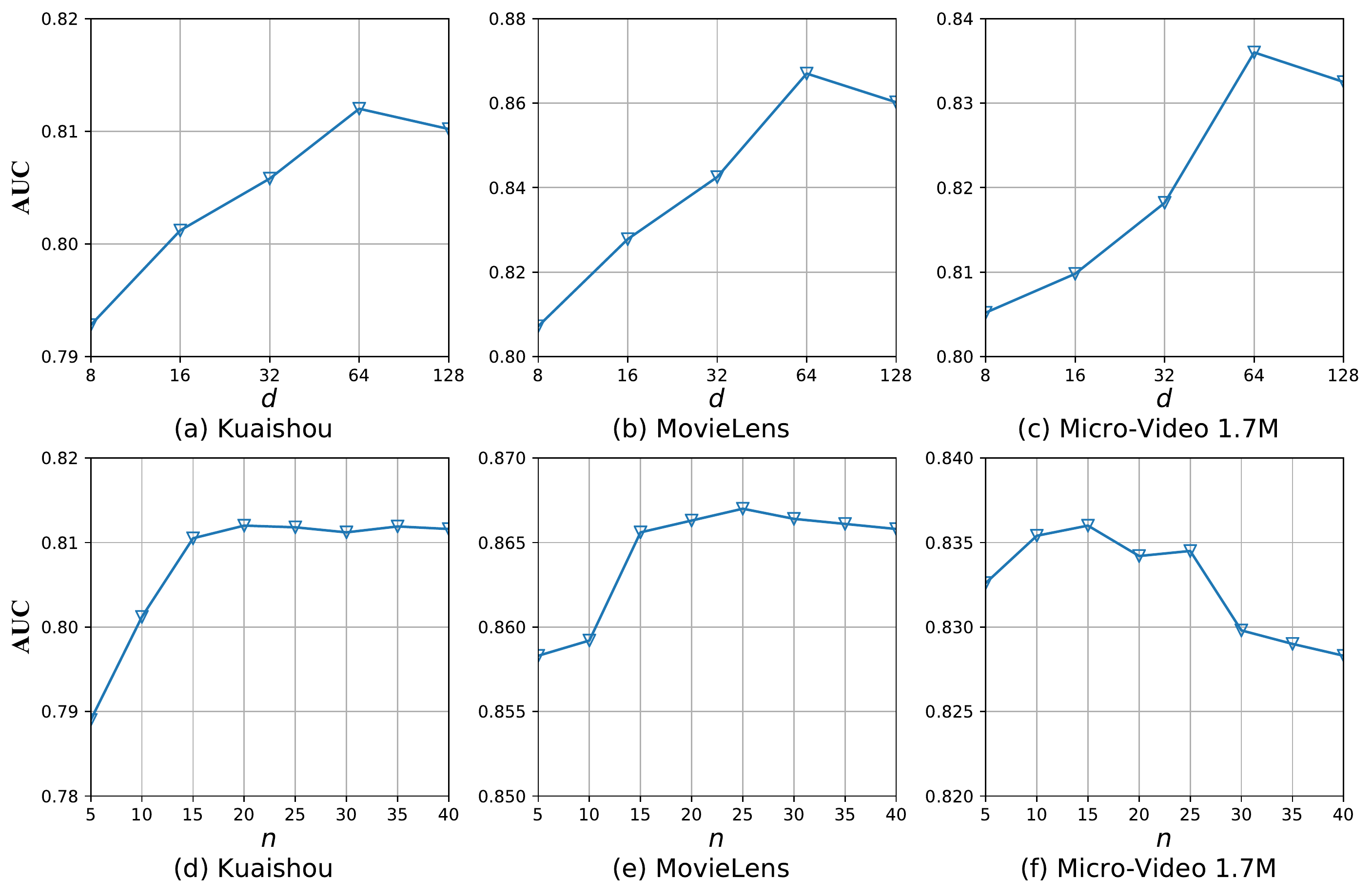}
  \vspace{-2em}
  \caption{Impact of embedding dimension (top row) and sampled neighbor size (bottom row)}
  \label{fig:sen}
  \vspace{-1em}
  \Description{}
\end{figure}

We study sensitivity of HyperCTR on the key hyperparameters using the three public datasets. The hyper-parameters play important roles in HGCN-based model, as they determine how the node embeddings are generated. We conduct experiments to analyze the impact of two key parameters which are the embedding dimension \(d\) and the size of sampled neighbors set for each node. According to Figure~\ref{fig:sen}, we can note that: 1) When \(d\) varies from 8 to 256, all evaluation metrics increase in general since better representations can be learned. However, the performance becomes stable or slightly worse when \(d\) further increases. This may due to over-fitting. 2) When the neighbor size varies from 5 to 40, all evaluation metrics increase at first as suitable amount of neighborhood information are considered. When the size of neighbors exceeds a certain value, performance decreases slowly which may due to irrelevant neighbors. The most ideal neighbor size is in the range of 15 to 25.

\subsubsection{Scalability Analysis}
\begin{figure}[h!]
  \centering
  \includegraphics[scale=0.3]{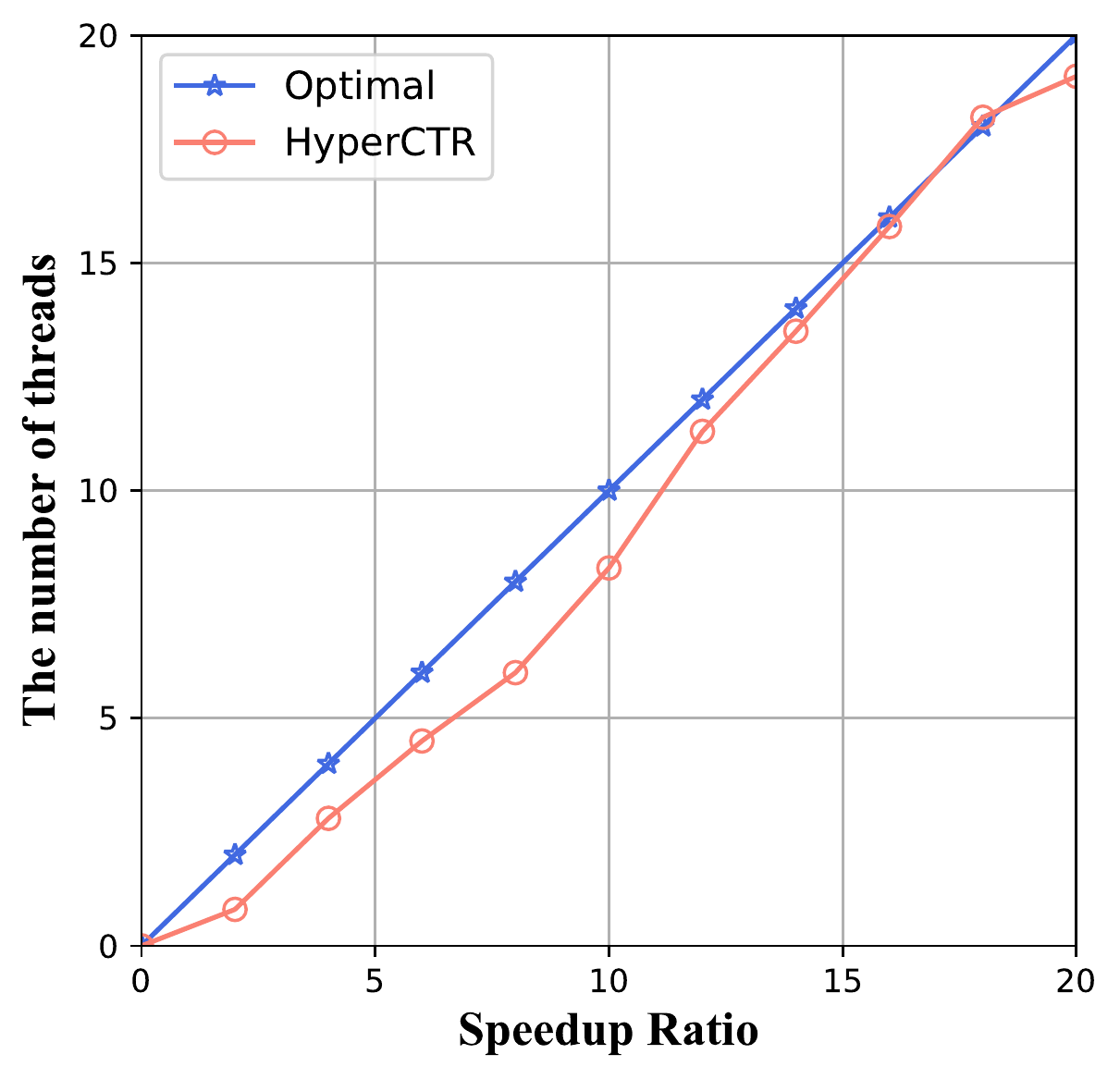}
  \vspace{-1em}
  \caption{Scalability of HyperCTR}
  \label{fig:scal}
  \vspace{-1em}
\end{figure}

As GCN-based networks are complex and contain such a large number of nodes in the real world application scenario, it is necessary for a model being feasible to be applied in the large-scale datasets. We investigate the scalability of HyperCTR model optimized by gradient descent, which deploys multiple threads for parallel model optimization. Our experiments are conducted in a computer server with 24 cores and 512GiB memory. We run experiments with different threads from 1 to 24. We depict in Figure~\ref{fig:scal} the speedup ratio vs. the number of threads. The speedup ratio is very close to linear, which indicates that the optimization algorithm of the HyperCTR is reasonably scalable.

\begin{figure}
  \centering
  \includegraphics[width=\linewidth]{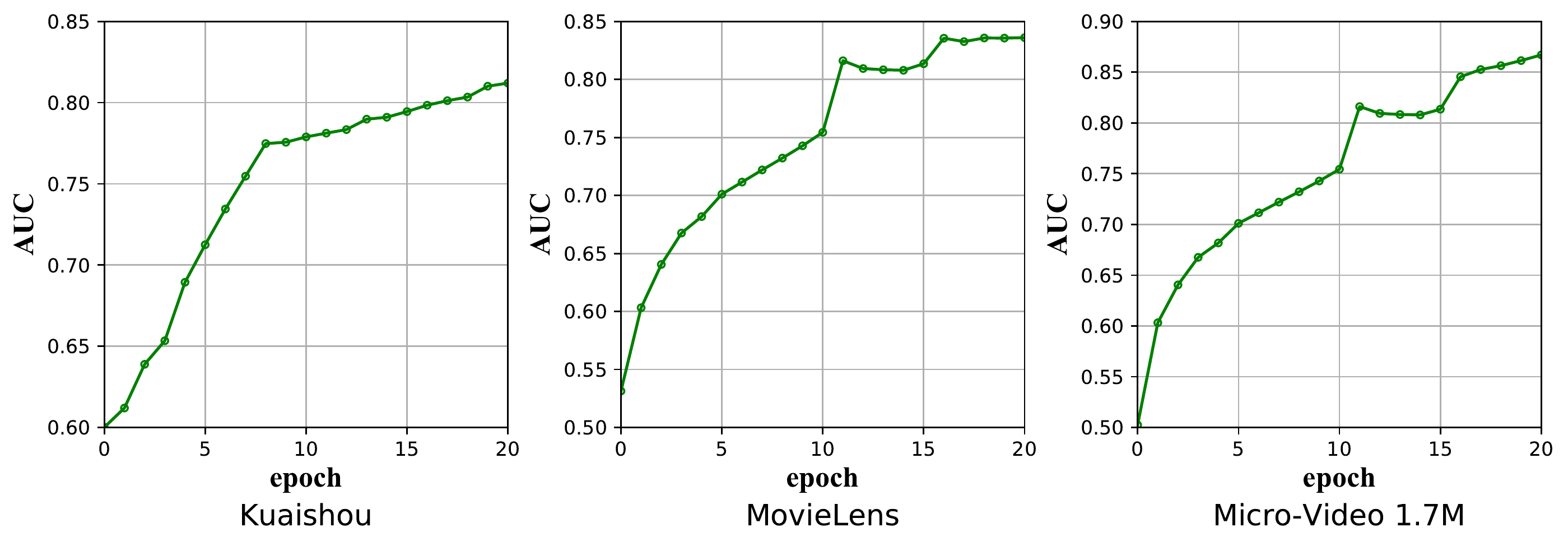}
  \vspace{-2em}
  \caption{Learning process of HyperCTR.}
  \label{fig:process}
  \vspace{-1.5em}
\end{figure}

\subsubsection{Model Training}
To depict our model training process, we plot the learning curves of HyperCTR, as shown in Figure~\ref{fig:process}. The three subfigures are the AUC curves of the multi-modal hypergraph framework when training on three datasets. Every epoch of the \(x\)-axis is corresponding to the iteration over 5\% of the training set.

\section{Related Work}
\textbf{CTR prediction} Learning the effect of feature interactions seems to be crucial for accurate CTR prediction. Factorization Machines (FMs)~\cite{blondel2016higher, 2010fm} are proposed to model pairwise feature interactions in terms of the vectors corresponding to the involved features. AutoFIS~\cite{Liu2020AutoFIS} and UBR4CTR~\cite{Qin2020} further improve FM by removing the redundant feature interactions and retrieving a limited number of historic behavior that are most useful for each CTR prediction target. However, a FM-based model considers learning shallow representation, and it thus is unable to model the features faithfully. Deep Neural Networks (DNNs) are exploited for CTR prediction in order to automatically learn feature representations and higher-order feature interactions. DSTN~\cite{2019DeepSpatio} integrates heterogeneous auxiliary data (i.e., contextual, clicked and unclicked ads) in a unified framework based on the DNN model. Further, the other stream of models focus more on mining temporal patterns from sequential user behavior. GRU4Rec~\cite{hidasi2016sessionbased} is based on RNN. It is the first work which uses the recurrent cell to model sequential user behavior. MIMN~\cite{Pi2019} applies the LSTM/GRU operations for modeling users' lifelong sequential behavior.

\textbf{Exploiting multi-modal representation}
Some works focus on the multi-modal representation in the area of multi-modal CTR prediction. Existing multi-modal representations have mostly been applied to recommender systems and have been grouped into two categories: joint representations and coordinated representations~\cite{Wei2019}. Joint representations usually combine the uni-modal information and project into the same representation space~\cite{cheng2019mmalfm,cheng2016effective,zhang2019neural,2021Reinforced,2021MetaPath}. Although, visual or textual data and are increasingly used in the multi-modal domain~\cite{Li2020}, they are suited for situations where all of the modalities are present during inference, which is hardly guaranteed in social platforms. Different from the joint representations, the coordinated models learn separate representations for each modality but coordinate them with constraints~\cite{Wei2019}. Since the modal-specific information is the factor for the differences in each modality signals, the model-specific features are inevitably discarded via similar constrains. In contrast, we introduce a novel model which respectively models the information augmentation and group-aware network problems to address the limitations in existing works.

\textbf{Graph Convolution Network}
Our proposed model uses the GCNs technique to represent the users and items, which has been popularly used for modeling the social media data. In ~\cite{hamilton2017inductive} the authors proposed a general inductive framework which leverages the content information to generate node representation for unseen data. In ~\cite{ying2018graph} the authors developed a large-scale deep recommendation engine on Pinterest for image recommendation. In their model, graph convolutions and random walks are combined to generate the representations of nodes. In ~\cite{berg2017graph} the authors proposed a graph auto-encoder framework based on message passing on the bipartite interaction graph. However, these methods cannot model the multi-modal data including cases where data correlation modeling is not straightforward~\cite{Feng2019}.

\section{Conclusion}
In this paper, we model temporal user preferences and multi-modal item attributes to enhance the accuracy of CTR prediction. We design a novel HGCN-based framework, named HyperCTR, to leverage information interaction between users and micro-videos by considering different modalities. We also refine user presentation from two aspects: time-aware and group-aware. With the stacking of hypergraph convolution networks, a self-attention and the fusion layer, our proposed model provides more accurate modeling of user preferences, leading to improved performance.

\section{Acknowledgments}
This work is supported in part by NSF under grants III-1763325, III-1909323,  III-2106758, and SaTC-1930941, and ARC under grants DP200101374 and LP170100891.


\bibliographystyle{ACM-Reference-Format}
\bibliography{cikm_fullpaper}
\balance
\end{document}